\documentclass[12pt]{article}
\usepackage{amsmath,amssymb,amsthm}
\numberwithin{equation}{section}
\usepackage{hyperref}
\usepackage{units}
\usepackage{color}
\usepackage[T1]{fontenc}
\usepackage[utf8]{inputenc}
\usepackage{authblk}
\usepackage{bm}
\usepackage{graphicx}

\usepackage{datetime}
\usepackage{color}

\usepackage[toc,page]{appendix}

\title{Quantization of the one-dimensional free harmonic oscillator as an example of the application of the node theorem and the MacDonald-Hylleraas-Undheim theorem\vspace{10mm}} 

\author[]{Kunle Adegoke\thanks{Corresponding author: adegoke00@gmail.com, kunle.adegoke@yandex.com}}
\author[]{Adenike Olatinwo}

\affil{Department of Physics and Engineering Physics, \mbox{Obafemi Awolowo University}, Ile-Ife, Nigeria}

\begin{document}

\date{}

\maketitle

\begin{abstract}
\noindent Using heuristic arguments alone, based on the properties of the wavefunctions, we obtain the energy eigenvalues and the corresponding eigenfunctions of the one-dimensional harmonic oscillator. This approach is considerably simpler and is perhaps more intuitive than the traditional methods of solving a differential equation and manipulating operators.
\end{abstract}

\section{Introduction}\label{sec.sj5w7hr}
As is well-known, in some fortunate cases, the foreknowledge of certain properties of the states of a quantum system, based on its symmetries, can facilitate finding the eigenvalues and eigenfunctions of the system for such states. In this note we demonstrate this fact by applying the node theorem and the MacDonald-Hylleraas-Undheim theorem to quantize the one-dimensional free harmonic oscillator, described by the Hamiltonian
\begin{equation}
H =  - \frac{{\hbar ^2 }}{{2m}}\frac{{d^2 }}{{dx^2 }} + \frac{1}{2}m\omega ^2 x^2,\quad-\infty<x<\infty\,,
\end{equation}
where $H$ lives in a Hilbert space $\mathcal H$ of real functions, with inner product defined for any pair of vectors $f(x)$ and $g(x)$ in $\mathcal H$ by
\[
(f(x),g(x)) = \int_{ - \infty }^\infty  {f(x)g(x)dx}\,. 
\]
The symmetry argument presented here is much easier than the traditional methods of solving the one-dimensional harmonic oscillator problem, discussed in every book on quantum mechanics, namely the direct solution of the time-independent Schr\"odinger equation and the operator method.

\bigskip

The node theorem for one-dimensional Hamiltonians states that the ground state of a system has no nodes (zeros {\em between} the boundaries) while the $r^{th}$ excited state has exactly $r$~nodes. This theorem is a direct consequence of the fact that the one dimensional Schr\"odinger equation is an example of a Sturm-Liouville equation (see reference~\cite{moriconi} and the references therein).

\bigskip

The MacDonald-Hylleraas-Undheim theorem~\cite{shull} (henceforth MHU theorem), first proved by Hylleraas and Undheim~\cite{hylleraas}, and later by MacDonald~\cite{macdonald}, states that:
\begin{quotation}
\noindent If the roots $\varepsilon_i$ of a secular equation are ordered such that \mbox{$\varepsilon_0\le \varepsilon_1\le \varepsilon_2\le\cdots\le \varepsilon_n$}, then each such root is an upper bound to the corresponding exact eigenvalue, $E_i$, that is $\varepsilon_i\ge E_i$.
\end{quotation}
\section{Matrix representation of $H$}
\subsection{Properties of the eigenfunctions of $H$}
Let $\varphi_1(x)$, $\varphi_2(x)$, $\ldots$ be the yet to be determined eigenfunctions of $H$, with corresponding eigenvalues $E_1, E_2, \ldots$. We make the following observations:
\begin{itemize}
\item The functions $\varphi_1(x), \varphi_2(x),\ldots$ are required to vanish at the boundaries, that is, $\varphi_r(\pm\infty)=0$ for $r=0,1,2,\ldots$
\item Since the potential $V(x)$ is an even function of $x$, the wavefunctions $\varphi_r(x)$, $r=0,1,\ldots$ are eigenstates of the parity operator so that each $\varphi_r$ is either an even function of~$x$ or an odd function of~$x$.
\item Since the potential $V(x)=m\omega^2x^2/2$ approaches infinity as $x$ approaches infinity, the eigenvalues of $H$ form a discrete unbounded sequence~\cite{moriconi}, the states $\varphi_0(x)$, $\varphi_1(x)$, $\varphi_2(x)$, $\ldots$ are non-degenerate, and, using the node theorem ($\varphi_0$ has $0$ nodes, $\varphi_1$ has $1$ node, $\varphi_2$ has $2$ nodes and so on) we can arrange the corresponding energy eigenvalues as \mbox{$E_0< E_1< E_2<\cdots $} where $E_0$ is the ground state energy, $E_1$ the first excited state energy and so on.
\item If we build a finite $(n+1)\times(n+1)$ matrix $H=(H_{ij})$ in an $n+1$ dimensional subspace of $\mathcal H$ and obtain the eigenvalues $\varepsilon_0, \varepsilon_1,\ldots,\varepsilon_n$ of the secular equation $|H-\varepsilon I|=0$, arranged such that \mbox{$\varepsilon_0<\varepsilon_1\ldots<\varepsilon_n$}, then according to the MHU theorem, we have $E_0\le\varepsilon_0$, $E_1\le\varepsilon_1$, $E_2\le\varepsilon_2$, $\ldots$, $E_n\le\varepsilon_n$.
\end{itemize}
\subsection{Choice of basis functions}
Consider the following complete set of functions
\begin{equation}\label{equ.pniljol}
\varphi_r(x)=A_re^{-\alpha x^2/2}H_r(x\sqrt\alpha),\quad r=0,1,2,\ldots\,,\alpha>0\,,
\end{equation}
vectors of $\mathcal H$, where $H_s(y)$ is the Hermite polynomial of degree $s$ in the variable $y$ and
\begin{equation}
A_s{}^2=\frac1{2^ss!}\sqrt{\frac\alpha\pi}\,.
\end{equation}
Using the orthogonality property of the Hermite polynomials with respect to the exponential weight function, we note that the $\varphi_r(x)$ given in~\eqref{equ.pniljol} are an orthonormal set in $\mathcal H$, so that $(\varphi_r,\varphi_s)=\delta_{rs}$ for $r,s\in \{0,1,2,\ldots\}$, for {\bf\em any} $\alpha>0$. We hasten to emphasize that here no assumptions are made on $\alpha$ other than that it is a positive constant.

\bigskip

Obviously, the orthonormal functions $\varphi_r$ given in~\eqref{equ.pniljol} vanish at \mbox{$x=\pm\infty$} and have definite parity for each $r$ since Hermite polynomials satisfy $H_s(-y)=(-1)^sH_s(y)$. Furthermore since $H_s(y)$ is a polynomial of degree $s$ in $y$ with $s$ distinct roots in $(-\infty,\infty)$, each $\varphi_s$ has $s$ nodes and is therefore suitable to represent the $s^{th}$ excited state. Thus, the functions $\varphi_r(x)$, in addition to being suitable choice as basis functions for giving a matrix representation for the Hamiltonian $H$, are themselves potential candidates for eigenfunctions of $H$.

\bigskip

We shall need the following recurrence relations for the Hermite polynomials:
\begin{equation}\label{equ.twaf466}
2yH_s(y)=H_{s+1}(y)+2sH_{s-1}(y)
\end{equation}
and
\begin{equation}\label{equ.faoxak9}
\frac d{dy}\left(e^{-y^2}H_s(y)\right)=-e^{-y^2}H_{s+1}(y)\,.
\end{equation}
\subsection{Matrix elements of $H$}
Writing $H(x)=T(x)+V(x)$ with
\[
T(x)=-\frac{\hbar^2}{2m}\frac{d^2}{dx^2},\quad V(x)=\frac12 m\omega^2x^2\,,
\]
we have
\begin{equation}\label{equ.xw3efvr}
\begin{split}
T_{rs}=(\varphi_r,T\varphi_s)&=-\frac{\hbar^2}{2m}\left(\varphi_r,\frac{d^2\varphi_s}{dx^2}\right)\\
&\quad=\frac{\hbar^2}{2m}\left(\frac{d\varphi_r}{dx},\frac{d\varphi_s}{dx}\right)\,.
\end{split}
\end{equation}
and
\begin{equation}\label{equ.fwux92y}
V_{rs}=(\varphi_r,V\varphi_s)=\frac12m\omega^2\left(x\varphi_r,x\varphi_s\right)\,.
\end{equation}
Writing~\eqref{equ.pniljol} as
\[
e^{ - \alpha x^2 /2} \varphi _r (x) = A_r e^{ - \alpha x^2 } H_r (x\sqrt \alpha  )
\]
and using~\eqref{equ.faoxak9}, it is easy to establish that
\begin{equation}\label{equ.d4o2bla}
\frac{d}{{dx}}\varphi _r (x) = \alpha x\varphi _r (x) - \left[ {2\alpha (r + 1)} \right]^{1/2} \varphi _{r + 1} (x)\,.
\end{equation}
From the definition~\eqref{equ.pniljol} of $\varphi_r$ and the recurrence relation~\eqref{equ.twaf466} we have
\begin{equation}\label{equ.iltxgm8}
x\varphi _r (x)\sqrt {2\alpha }  = \varphi _{r + 1} (x)\sqrt {r + 1}  + \varphi _{r - 1} (x)\sqrt r
\end{equation}
Using~\eqref{equ.iltxgm8} in~\eqref{equ.d4o2bla} we get
\begin{equation}\label{equ.j6157e2}
\frac{d}{{dx}}\varphi _r (x) = \frac{{\sqrt {2\alpha } }}{2}\left( {\varphi _{r - 1} (x)\sqrt r  - \varphi _{r + 1} (x)\sqrt {r + 1} } \right)\,.
\end{equation}
Using~\eqref{equ.j6157e2} in~\eqref{equ.xw3efvr}, together with the orthonormalization condition \mbox{$(\varphi_r,\varphi_s)=\delta_{rs}$} we finally have
\begin{equation}\label{equ.d2706j8}
\begin{split}
T_{rs}  &= \frac{{\alpha \hbar ^2 }}{{4m}}\left[ {\sqrt {rs}\, \delta _{rs}  - \sqrt {r(s + 1)}\, \delta _{r,s + 2} } \right.\\
&\qquad\qquad\left. {\sqrt {s(r + 1)}\, \delta _{s,r + 2}  + \sqrt {(r + 1)(s + 1)}\, \delta _{rs} } \right]\,.
\end{split}
\end{equation}
We see that
\begin{equation}
\begin{split}
T_{rr} &= \frac{{\alpha \hbar ^2 }}{{4m}}(2r + 1),\quad T_{r,r + 2}  =  - \frac{{\alpha \hbar ^2 }}{{4m}}\sqrt {(r + 1)(r + 2)}  = T_{r + 2,r}\,,\\
T_{rs}&=0 \mbox{ if $|r-s|>2$}\,.
\end{split}
\end{equation}
Using~\eqref{equ.iltxgm8} in~\eqref{equ.fwux92y} we obtain
\begin{equation}\label{equ.wgkgb1x}
\begin{split}
V_{rs}  &= \frac{{m\omega^2 }}{{4\alpha}}\left[ {\sqrt {rs}\, \delta _{rs}  + \sqrt {r(s + 1)}\, \delta _{r,s + 2} } \right.\\
&\qquad\qquad\left. {\sqrt {s(r + 1)}\, \delta _{s,r + 2}  + \sqrt {(r + 1)(s + 1)}\, \delta _{rs} } \right]\,,
\end{split}
\end{equation}
so that
\begin{equation}
\begin{split}
V_{rr} &= \frac{m\omega^2}{{4\alpha}}(2r + 1),\quad V_{r,r + 2}  =  \frac{{m \omega^2 }}{{4\alpha}}\sqrt {(r + 1)(r + 2)}  = V_{r + 2,r}\,,\\
V_{rs}&=0 \mbox{ if $|r-s|>2$}\,.
\end{split}
\end{equation}
Finally, from~\eqref{equ.d2706j8} and \eqref{equ.wgkgb1x} we obtain the matrix representation of the one dimensional harmonic oscillator Hamiltonian $H$ as
\begin{equation}\label{equ.gujrrym}
\begin{split}
H_{rs}  &= \delta _{rs} \left( {\frac{{\alpha \hbar ^2 }}{{4m}} + \frac{{m\omega ^2 }}{{4\alpha }}} \right)(2r + 1)\\
&\qquad + \delta _{r + 2,s} \left( { - \frac{{\alpha \hbar ^2 }}{{4m}} + \frac{{m\omega ^2 }}{{4\alpha }}} \right)\sqrt {(r + 1)(r + 2)}\\ 
&\qquad\quad + \delta _{s + 2,r} \left( { - \frac{{\alpha \hbar ^2 }}{{4m}} + \frac{{m\omega ^2 }}{{4\alpha }}} \right)\sqrt {(r - 1)r}\,. 
\end{split}
\end{equation}
\section{Diagonalization of $H$}
With the matrix elements given in~\eqref{equ.gujrrym}, the problem of quantization of the one dimensional harmonic oscillator reduces to that of building finite matrices and finding the eigenvalues and eigenvectors of the matrices. Using this approach, convergence and accuracy of the results generally depend on making a judicious choice of the parameter $\alpha$. To this end, the variation principle is usually employed to make an optimum choice of $\alpha$. In this present study, it turns out that, in fact, there is a choice of $\alpha$ for which the $H$~matrix is exactly diagonal.

\bigskip

We see from the matrix elements of $H$ in~\eqref{equ.gujrrym} that if
\begin{equation}\label{equ.wh6p63u}
 - \frac{{\alpha \hbar ^2 }}{{4m}} + \frac{{m\omega ^2 }}{{4\alpha }}=0\,,
\end{equation}
then the Hamiltonian matrix $H$ becomes diagonal. Thus from~\eqref{equ.wh6p63u} the value of $\alpha$ for $H$ to be diagonal is fixed at
\begin{equation}
\alpha=\frac{m\omega}\hbar\,,
\end{equation}
and the eigenvalues of $H$ are then given by
\begin{equation}
\varepsilon_r=H_{rr}=\frac{\hbar\omega}2(2r+1)\,,\quad r=0,1,2,\ldots
\end{equation}
Since $H$ is diagonal, the eigenvalues are exact and the prescription of the MacDonald-Hylleraas-Undheim theorem in the last paragraph of section~\ref{sec.sj5w7hr} gives $E_0=\varepsilon_0$, $E_1=\varepsilon_1$, $E_2=\varepsilon_2$, $\ldots$., that is
\begin{equation}
E_r=\frac{\hbar\omega}2(2r+1)\,,\quad r=0,1,2,\ldots
\end{equation}
Now substituting $\alpha=m\omega/\hbar$ into~\eqref{equ.pniljol}, we have the corresponding energy eigenfunctions to be given by
\begin{equation}
\varphi _r (x) = \frac{1}{{\sqrt {2^r r!} }}\left( {\frac{{m\omega }}{{\pi \hbar }}} \right)^{1/4} H_r \left( {x\sqrt {\frac{{m\omega }}{\hbar }} }\, \right)\exp \left( { - \frac{{m\omega }}{{2\hbar }}x^2 } \right),\quad r = 0,1,2, \ldots
\end{equation}
\section{Concluding remarks}
Although the MacDonald-Hylleraas-Undheim theorem and the node theorem always work for one dimensional problems, it is not always that one gets lucky and is able to make a choice of a wavefunction parameter for which the Hamiltonian matrix is diagonal. Consider the one dimensional pure quartic oscillator, described by the Hamiltonian 
\[
H_q =  - \frac{{\hbar ^2 }}{{2m}}\frac{{d^2 }}{{dx^2 }} + \lambda x^4,\quad\lambda>0\,.  
\]
Using the same $\varphi_r(x)$ given in~\eqref{equ.pniljol} as basis functions, the matrix elements of $H_q$ are
\[
\begin{split}
H_{q _{rs}}  &= \delta _{rs} \left( {\frac{{\alpha \hbar ^2 }}{{4m}}(2r + 1) + \frac{{3\lambda }}{{4\alpha ^2 }}(2r^2  + 2r + 1)} \right)\\
&\quad + \delta _{r + 2,s} \left( { - \frac{{\alpha \hbar ^2 }}{{4m}} + \frac{{(2r + 3)\lambda }}{{2\alpha ^2 }}} \right)\sqrt {(r + 1)(r + 2)}\\ 
&\qquad + \delta _{s + 2,r} \left( { - \frac{{\alpha \hbar ^2 }}{{4m}} + \frac{{(2r - 1)\lambda }}{{2\alpha ^2 }}} \right)\sqrt {(r - 1)r}\\ 
&\quad\qquad + \delta _{r + 4,s} \frac{\lambda }{{4\alpha ^2 }}\sqrt {(r - 1)r(r + 5)(r + 6)}\\ 
&\qquad\qquad + \delta _{r + 4,s} \frac{\lambda }{{4\alpha ^2 }}\sqrt {(r - 5)(r - 4)(r + 1)(r + 2)}\,. 
\end{split}
\]
While the eigenvalues and eigenvectors of the finite dimension $H_q$ matrix can be calculated for any value of $\alpha$, it is obvious that no choice of $\alpha$ that is independent of the indices $r$ and $s$ can bring the $H_q$ matrix to a diagonal form, so that only approximate eigenvalues and eigenvectors can be obtained, with the accuracy depending on the choice of $\alpha$.

\end{document}